\documentclass{elsart}
\usepackage{amsmath,amssymb,curves}
\journal{Physics Letters B}

\newcommand{\del}{\partial}
\newcommand{\run}[1]{\tilde{\alpha}_{#1}}

\begin{document}
\begin{frontmatter}

\title{Running couplings for the simultaneous decoupling of heavy quarks}

\author[Austria]{Steven D.\ Bass\thanksref{ASF}},
\author[Australia]{R.J.\ Crewther},
\author[Brazil1,Brazil2]{F.M.\ Steffens\thanksref{FAPESP}},
\author[USA]{A.W.\ Thomas\thanksref{DOE}}

\address[Austria]{Institute for Theoretical Physics, 
Universit\"at Innsbruck,
Technikerstrasse 25, A~6020 Innsbruck, Austria}
\address[Australia]{Department of Physics and Centre 
for the Subatomic Structure of Matter (CSSM), 
University of Adelaide, SA 5005, Australia}
\address[Brazil1]{Instituto de Fisica Teorica, UNESP,
Rua Pamplona 145, 01405-900 S\~ao Paulo, SP, Brazil}
\address[Brazil2]{Mackenzie Presbiteriana University, FCBEE,
Rua da Consola{\c{c}}\~ao 930, 01302-907 S\~ao Paulo, SP, Brazil}
\address[USA]{Jefferson National Laboratory, 12000 Jefferson Avenue,
Newport News, VA~23606, U.S.A.}  

\thanks[ASF]{Supported by the Austrian Science Fund, contract P17778-N08.}
\thanks[FAPESP]{Supported by FAPESP (03/10754-0) and CNPq (308932/2003-0).}
\thanks[DOE]{Supported by DOE contract DE-AC05-84ER40150, under which 
SURA operates Jefferson National Laboratory.}

\begin{abstract}
Scale-invariant running couplings are constructed for several quarks
being decoupled together, \emph{without} reference to intermediate
thresholds.  Large-momentum scales can also be included. The result 
is a multi-scale generalization of the renormalization group applicable 
to any order.  Inconsistencies in the usual decoupling procedure with 
a single running coupling can then be avoided, e.g.\ when cancelling
anomalous corrections from $t,b$ quarks to the axial charge of the 
proton. 
\end{abstract}


\end{frontmatter}

Normally, heavy quarks are decoupled one at a time, in order of decreasing 
mass $m_q$, $q = t,b, \ldots$.  First the top quark is removed, with
the $b$- and $c$-quark thresholds left intact.  Then $b$ is decoupled
without destroying the $c$-quark threshold, and so on:
\begin{equation} m_t \to \infty\ \mbox{first},\
\mbox{then}\  m_b \to \infty, \
\mbox{then}\  m_c \to \infty.
\label{eq1}
\end{equation}
This sequential approach has two basic problems:
\begin{enumerate}
\item On a logarithmic scale, $m_t$ may be sufficiently large compared
      with $m_b$, but not\footnote%
{The $t$ quark, the $W^\pm$ and $Z^0$ bosons, and (presumably) the Higgs 
boson, provide another example of this problem. When more than one of these 
particles sets the scale, or there are momenta of similar magnitude, 
it is not reasonable to suppose that one of the corresponding 
logarithms dominates the others. See \cite{smir,blum,been,denn,brod}
for alternative methods.}
$m_b$ compared with $m_c$.
\item More seriously, it is inconsistent to retain both 
      $O(\bar{\alpha}(m_t))$ and $O(\bar{\alpha}(m_{b,c}))$ corrections
      when the first step $m_t \to \infty$ assumes
      $\bar{\alpha}(m_t) \ll \bar{\alpha}(m_{b,c})$; here 
      $\bar{\alpha}(M)$ is the usual running coupling at scale $M$.
\end{enumerate}

Our proposal is to replace the standard procedure by a new technique
based on an extended renormalization group (RG) where heavy-quark 
decoupling is \emph{genuinely} simultaneous and all mass logarithms are 
summed together.  We show the need for this with an explicit example: 
anomaly cancellation for $t,b$ quarks decoupled from the axial charge
of the proton.

We let two or more mass logarithms grow large together relative to a 
light-quark scale $\bar{\mu}$, e.g.\ 
\begin{equation}
\ln(m_c/\bar{\mu}) \lesssim \ln(m_b/\bar{\mu}) \lesssim \ln(m_t/\bar{\mu})\,
\to\, \mbox{large}  
\label{eq2}
\end{equation}
for three heavy quarks. These heavy quarks are decoupled together, 
i.e.\ \emph{in a single step, without reference to intermediate thresholds}.
Instead of a single running coupling $\bar{\alpha}(M)$, there are separate
running couplings $\alpha_t, \alpha_b, \ldots$, one for each heavy quark 
$h = t,b, ...$. The leading asymptotic power of an amplitude 
$\mathcal{A}$ is expanded as a \emph{multinomial} in 
$\alpha_t,\alpha_b, ...$,
\begin{equation}
\mathcal{A} = \sum_{k\ell\ldots}\alpha_t^k\alpha_b^\ell\ldots
                   \mathcal{A}_{k\ell\ldots}
\label{a} 
\end{equation}
with all $\{m_h\}$ dependence contained in $\alpha_t,\alpha_b, ...$.
The coefficients $\mathcal{A}_{k\ell\ldots}$ are
amplitudes produced by the residual light-quark theory.  They can be
identified by observing that the special case
\begin{equation}
\ln(m_t/\bar{\mu}) \gg \ln(m_b/\bar{\mu}) \gg \ldots
\label{a1} 
\end{equation}
of the simultaneous limit (\ref{eq2}) must give the same results as 
a direct application of the sequential limit (\ref{eq1}).  This technique 
depends on the fact that the residual theory is the \emph{same}, 
irrespective of how the heavy quarks are decoupled.

A key step is the construction of simultaneous running couplings 
$\alpha_t,\alpha_b,\ldots$ for the general limit (\ref{eq2}).
Each $\alpha_h$ is RG invariant, depends on \emph{all} of the 
mass logarithms in Eq.~(\ref{eq2}), and in the limit (\ref{a1}), 
can be expanded in running couplings $\run{t}$, $\run{b}$, and 
(if needed) $\run{c}$ associated with separate steps in the 
sequential limit (\ref{eq1}).

Let $\alpha_f = g_f^2/4\pi$ and $m_{qf}$ be the strong coupling 
and quark masses renormalized in the presence of $f$ flavours and
three colours according to the $\overline{\mbox{\small MS}}$  
(modified minimal subtraction) scheme.  We need the functions
\begin{align}
\beta_f(x) &= - \frac{x^2}{6\pi}(33 - 2f)
    - \frac{x^3}{12\pi^2}(153 - 19f) + O\bigl(x^4\bigr) \ , \\[1mm]
\delta_f(x) &= - \frac{2x}{\pi} + O\bigl(x^2\bigr) 
\label{a2}
\end{align}
associated with the Callan-Symanzik operator
\begin{equation}
\mathcal{D}_f = \bar{\mu}\frac{\del\ }{\del\bar{\mu}}
      + \beta_f(\alpha_f)\frac{\del\ \ }{\del\alpha_f}
        + \delta_f(\alpha_f)\sum_{q=1}^f m_{qf}
             \frac{\del\,\ \ \ }{\del m_{qf}}\ ,
\end{equation}
where $\bar{\mu} = \sqrt{4\pi}e^{-\gamma/2}\mu_{\mathrm{dim}}$ is 
the $\overline{\mbox{\small MS}}$ scale ($\mu_{\mathrm{dim}}$ being 
the scale for dimensional renormalization). Note that 
$\bar{\mu}$ is chosen to be independent of $f$, that it sets the 
fixed low-energy scale needed to define simultaneous limits, as in 
Eq.~(\ref{eq2}), and that the heavy-quark masses for these 
limits are renormalized in the \emph{six}-flavour theory:
\begin{equation}
m_h \equiv m_{h6} \quad , \quad h = t,b,\ldots
\end{equation}

The running coupling $\alpha_t$ for the \emph{heaviest} quark $t$ is 
defined by adapting Witten's prescription \cite{witten} for a single
heavy quark.  We set $f=6$,
\begin{equation}
\ln(m_t/\bar{\mu})
= \int^{\alpha_t}_{\alpha_6}\!\!dx\,\{1-\delta_6(x)\}/\beta_6(x)
\label{d}
\end{equation}
and require that $\alpha_6$ be matched \emph{directly} to the 
coupling $\alpha_\mathsf{f}$ of the residual $\mathsf{f}$-flavour theory 
in the limit (\ref{eq2}).  By construction, $\alpha_t$ is scale invariant
with respect to the original $f=6$ theory:
\begin{equation}
\mathcal{D}_6\alpha_t = 0 \ .
\end{equation}

Scale-invariant definitions of the other running couplings are then
easily produced.  The idea is to exploit a characteristic feature of 
simultaneous decoupling --- the presence of large \emph{scale-invariant} 
logarithms such as\footnote%
{In the standard procedure (\ref{eq1}), one chooses a scale for each 
step such that large logarithms never appear \cite{CWZ}.  For simultaneous 
limits (\ref{eq2}), that is no longer possible.\label{scale}}
\begin{equation} 
\ln(m_t/m_b) = \ln(m_t/\bar{\mu}) - \ln(m_b/\bar{\mu}) \ . 
\end{equation}
This logarithm can be used to set the relative scale between 
$\alpha_t$ and $\alpha_b$,
\begin{equation} 
 \ln\bigl(m_t/m_b\bigr)_6
= \int^{\alpha_t}_{\alpha_b}\!\!dx\,\{1-\delta_5(x)\}/\beta_5(x) \ ,
\label{g}
\end{equation}
with $\alpha_c$ defined similarly if the $c$ quark is also 
being decoupled:\footnote%
{These definitions make sense for \emph{all} mass values consistent 
with $m_t \geqslant m_b \geqslant m_c$. This includes equal-mass cases, 
where running couplings coincide, e.g.\ $\alpha_t = \alpha_b$ for 
$m_t = m_b$. \protect\label{equal} }
\begin{equation} 
 \ln\bigl(m_b/m_c\bigr)_6
= \int^{\alpha_b}_{\alpha_c}\!\!dx\,\{1-\delta_4(x)\}/\beta_4(x) \ .
\label{h}
\end{equation}
Note that $\alpha_b$ and $\alpha_c$ exhibit the full $f=6$ scale
invariance of the original theory
\begin{equation}
\mathcal{D}_6\alpha_b = 0 = \mathcal{D}_6\alpha_c 
\end{equation}
\emph{irrespective} of the integrands chosen in Eqs.\ (\ref{g}) and 
(\ref{h}).

We will show that these definitions produce simultaneous running couplings
$\alpha_t, \alpha_b,\ldots$ with the property
\begin{equation}
\alpha_h 
\underset{\mathrm{seq}}{=} \run{h} + O\bigl(\run{t,b,\ldots}^3\bigr)
\label{i1}
\end{equation}
in the sequential limit (\ref{a1}), where $\run{t},\run{b},\ldots$ are
Witten's running couplings for successive steps in Eq.~(\ref{eq1}):
\begin{align}
\ln\bigl(m_{t6}/\bar{\mu}\bigr)
&= \int^{\run{t}}_{\alpha_6}\!\!dx\,\{1-\delta_6(x)\}/\beta_6(x) \ ,
\label{l1} \\
\ln\bigl(m_{b5}/\bar{\mu}\bigr)
&= \int^{\run{b}}_{\alpha_5}\!\!dx\,\{1-\delta_5(x)\}/\beta_5(x) \ ,
\label{l2} \\
\ln\bigl(m_{c4}/\bar{\mu}\bigr)
&= \int^{\run{b}}_{\alpha_4}\!\!dx\,\{1-\delta_4(x)\}/\beta_4(x) \ .
\label{l3}
\end{align}
Eq.~(\ref{i1}) says that $\alpha_h$ and $\run{h}$ coincide asymptotically
in leading order (LO) and next-to-leading order (NLO)%
\footnote{For $\alpha_t$ and $\run{t}$, the agreement is to all orders
in the leading power: the defining equations (\ref{d}) 
and (\ref{l1}) coincide.\label{same}}.
LO agreement ensures that any amplitude $\mathcal{A}$ 
which can be expanded in the sequential limit is \emph{analytic}
in $\{\alpha_h\}$ when expanded in the simultaneous limit, as
foreshadowed in Eq.~(\ref{a}), while agreement to NLO accuracy is 
convenient for practical purposes.

For example, let us decouple $t,b$ simultaneously from an amplitude 
$\mathcal{A}$. We match the simultaneous expansion (\ref{a}) of 
$\mathcal{A}$ to terms produced by sequential decoupling, where first 
the top quark is decoupled with $m_b$ held fixed,
\begin{equation}
\mathcal{A} 
 = \mathcal{A}_0 + \run{t}\mathcal{A}_1 + \run{t}^2\mathcal{A}_2  
   + O(\run{t}^3) 
\label{m1}
\end{equation}
and then $b$ is decoupled from each of the residual five-flavour 
coefficient amplitudes 
$\mathcal{A}_k$, $k = 0,1,2,\ldots$ to produce four-flavour amplitudes 
$\mathcal{A}_{k\ell}$:
\begin{equation}
\mathcal{A}_k = \mathcal{A}_{k0} + \run{b}\mathcal{A}_{k1} 
                + \run{b}^2\mathcal{A}_{k2}  + O(\run{b}^3) \ .
\label{m2}
\end{equation}
Combine Eqs.~(\ref{m1}) and (\ref{m2}), collecting terms of the same
power in $\run{t,b}$:
\begin{equation} 
\mathcal{A} \underset{\mathrm{seq}}{=} \mathcal{A}_{00} 
 + \bigl\{\run{t}\mathcal{A}_{10} + \run{b}\mathcal{A}_{01}\bigr\} 
 + \bigl\{\run{t}^2\mathcal{A}_{20} + \run{t}\run{b}\mathcal{A}_{11} 
          + \run{b}^2\mathcal{A}_{02}\bigr\} + O(\run{t,b}^3)\ . 
\label{n1}
\end{equation}
Then, for \emph{any} fixed value of $\ln(m_t/\bar{\mu})/\ln(m_b/\bar{\mu})$
from 1 to infinity, the linear (LO) and quadratic (NLO) corrections for 
\emph{simultaneous} decoupling involve the \emph{same} coefficient 
amplitudes as in Eq.~(\ref{n1}), because of the property (\ref{i1}):  
\begin{equation} 
\mathcal{A}  \underset{\mathrm{sim}}{=} \mathcal{A}_{00} 
 + \bigl\{\alpha_t \mathcal{A}_{10} + \alpha_b \mathcal{A}_{01}\bigr\} 
 + \bigl\{\alpha_t^2 \mathcal{A}_{20} + \alpha_t\alpha_b\mathcal{A}_{11} 
          + \alpha_b^2 \mathcal{A}_{02}\bigr\} + O(\alpha_{t,b}^3)\ . 
\label{n2}
\end{equation}
Coefficient amplitudes for higher orders can be calculated directly
from cubic and higher-order corrections to Eqs.~(\ref{i1}) and (\ref{n1}).
Phenomenology based on \emph{single}-limit expansions like (\ref{n2}) is 
perfectly consistent because all running couplings are of the same
order, e.g.\ $\alpha_t \sim \alpha_b$.

The proof of Eq.~(\ref{i1}) depends on RG equations which relate the 
initial $F$-flavour and residual $f$-flavour theories. These equations
are equivalent to the rule \cite{BCST2} $\mathcal{D}_F = \mathcal{D}_f$ 
when these operators act on $f$-flavour quantities.  Applied to 
$\alpha_f$ and $\ln m_{qf}$, this rule yields Witten's 
relations \cite{witten}, generalized to include simultaneous cases 
(\ref{eq2}):
\begin{equation}
\mathcal{D}_F\alpha_f = \beta_f(\alpha_f) \ , \quad
\mathcal{D}_F \ln m_{qf} = \delta_f(\alpha_f)  \quad
\mbox{for all } f < F.
\label{o1}
\end{equation}
An immediate consequence of Eq.~(\ref{o1}) is that the sequential
running couplings defined by Eqs.~(\ref{l1})--(\ref{l3}) are all
RG$_{F=6}$ invariant \cite{BCST2}:
\begin{equation}
\mathcal{D}_6\run{t} = \mathcal{D}_6\run{b} = \mathcal{D}_6\run{c} 
 = 0 \ .
\end{equation}

The proof also depends on matching conditions 
\cite{BCST2,BW,larin,CKS,rodrigo} 
for the ${f=4,5,6}$ couplings and masses in Eqs.~(\ref{l1})--(\ref{l3}). 
Let the decoupling of $h$ correspond to reducing the number of flavours 
from $f+1$ to $f$. Then a scale-invariant \emph{matching function} 
$\mathcal{F}_{f+1 \to f}$ can be defined by the condition \cite{BCST2}
\begin{equation}
\int_{\alpha_{f+1}}^{\run{h}}\! dx/\beta_{f+1}(x)
= \int_{\alpha_f}^{\run{h}}\! dx/\beta_f(x)
  + \mathcal{F}_{f+1 \to f}(\run{h}) \ .
\label{p1}
\end{equation}
It contributes at NNLO (next-to-NLO):
\begin{equation}
\mathcal{F}_{f+1 \to f}(\run{h})
= - \frac{11}{12\pi(33 - 2f)}\run{h} + O\bigl(\run{h}^2\bigr) \ .
\label{p2}
\end{equation}
There is also an invariant mass-matching function 
$\mathcal{G}_{f+1 \to f}$,
\begin{equation}
\ln\frac{m_{q(f+1)}}{m_{qf}}
=  \int_{\alpha_f}^{\run{h}}\! dx \frac{\delta_f(x)}{\beta_f(x)}
 - \int_{\alpha_{f+1}}^{\run{h}}\! dx 
         \frac{\delta_{f+1}(x)}{\beta_{f+1}(x)}
 + \mathcal{G}_{f+1 \to f}(\run{h})
\label{p3}
\end{equation}
which contributes at NNNLO (next-to-NNLO) \cite{BCST2}:
\begin{equation}
\mathcal{G}_{f+1 \to f}(\run{h})
= - \frac{89}{432\pi^2}\run{h}^2 + O\bigl(\run{h}^3\bigr) \ .
\label{p4}
\end{equation}

Now let Eqs.~(\ref{g}), (\ref{l1}), (\ref{l2}) and the identity
\begin{equation}
\ln\bigl(m_t/m_b\bigr)_6
= \ln(m_{t6}/\bar{\mu}) - \ln(m_{b5}/\bar{\mu}) 
  - \ln(m_{b6}/m_{b5})
\end{equation}
be combined with Eqs.~(\ref{p1}) and (\ref{p3}) for the case 
$f=5,\ h=t$.  The result is a relation between the various running
couplings for $t$ and $b$:
\begin{equation}
\int_{\run{t}}^{\alpha_t}\!dx\frac{1 - \delta_5(x)}{\beta_5(x)}
\underset{\mathrm{seq}}{=} 
\int_{\run{b}}^{\alpha_b}\!dx\frac{1 - \delta_5(x)}{\beta_5(x)}
  + \mathcal{F}_{6 \to 5}(\run{t}) - \mathcal{G}_{6 \to 5}(\run{t}) \ .
\label{q1}
\end{equation}
The $b,c$ running couplings are similarly related:
\begin{equation}
\int_{\run{b}}^{\alpha_b}\!dx\frac{1 - \delta_4(x)}{\beta_4(x)}
\underset{\mathrm{seq}}{=} 
\int_{\run{c}}^{\alpha_c}\!dx\frac{1 - \delta_4(x)}{\beta_4(x)}
  + \mathcal{F}_{5 \to 4}(\run{b}) - \mathcal{G}_{5 \to 4}(\run{b}) \ .
\label{q2}
\end{equation}
Eq.~(\ref{i1}) is an immediate consequence of Eqs.~(\ref{q1}) and 
(\ref{q2}): the matching functions do not contribute at NLO, and we 
have\footnotemark[0\ref{same}] 
$\alpha_t \underset{\mathrm{seq}}{=} \run{t}$
for the top couplings. Indeed, the NNLO corrections are easily found:
\begin{align}
\alpha_b &\underset{\mathrm{seq}}{=} 
  \run{b} - \frac{11}{72\pi^2}\run{b}^2\run{t} 
             + O\bigl(\run{t,b}^4\bigr) \ , \label{q3} \\[1mm]
\alpha_c &\underset{\mathrm{seq}}{=} 
  \run{c} - \frac{11}{72\pi^2}\run{c}^2(\run{b} + \run{t}) 
             + O\bigl(\run{t,b,c}^4\bigr) \ .
\end{align}

As noted below Eq.~(\ref{d}), $\alpha_6$ and $m_{q6}$ have to be 
matched \emph{directly} to their counterparts in the residual
$\mathsf{f}$-flavour theory.  This is done in two steps:  we 
replace Witten's invariant mass \cite{witten} $\tilde{m}_h$ for 
the heavy quark $h$ in Eq.~(\ref{p1})
\begin{equation}
\ln(\tilde{m}_h/\bar{\mu}) 
= \int_{\alpha_{f+1}}^{\run{h}}\!dx/\beta_{f+1}(x)
\end{equation}
with a set of \emph{simultaneous} RG$_{F=6}$ invariant masses 
$\overline{m}_t, \overline{m}_b, \ldots$, and then define simultaneous 
matching functions $\mathcal{F}_{6 \to \mathsf{f}}$ and 
$\mathcal{G}_{6 \to \mathsf{f}}$.

In the case of simultaneous $t,b$ decoupling, where the residual theory 
has four flavours, $\overline{m}_t$ and $\overline{m}_b$ are defined 
by the equations
\begin{align}
\ln(\overline{m}_t/\bar{\mu}) 
&= \int_{\alpha_6}^{\alpha_t}\!dx/\beta_6(x) \ , \nonumber  \\
\ln(\overline{m}_t/\overline{m}_b)
&= \int_{\alpha_b}^{\alpha_t}\!dx/\beta_5(x) \ .
\label{r1}
\end{align}
Then $\mathcal{F}_{6 \to 4}$ is defined by the condition
\begin{equation}
\ln(\overline{m}_b/\bar{\mu}) 
 = \int_{\alpha_4}^{\alpha_b}\!dx/\beta_4(x) 
   + \mathcal{F}_{6 \to 4}(\alpha_t, \alpha_b) \ .
\label{r2}
\end{equation}
Being RG$_{F=6}$ invariant, $\mathcal{F}_{6 \to 4}$ can depend only
on $\alpha_t$ and $\alpha_b$.  The matching condition between $\alpha_6$
and $\alpha_4$ for simultaneous $t,b$ decoupling is the result of 
eliminating $\alpha_t$ and $\alpha_b$ from Eqs.~(\ref{r1}) and (\ref{r2}).

An alternative definition of $\mathcal{F}_{6 \to 4}$ is obtained by 
eliminating $\overline{m}_t$ and  $\overline{m}_b$ from Eqs.~(\ref{r1}) 
and (\ref{r2}):
\begin{equation}
\mathcal{F}_{6 \to 4}(\alpha_t,\alpha_b) 
 = - \int_{\alpha_4}^{\alpha_b}\!dx \frac{1}{\beta_4(x)}
 - \int_{\alpha_b}^{\alpha_t}\!dx \frac{1}{\beta_5(x)}
 - \int_{\alpha_t}^{\alpha_6}\!dx \frac{1}{\beta_6(x)} \ .
\label{r3}
\end{equation}
Comparison with Eq.~(\ref{p3}) then suggests the following RG$_{F=6}$
invariant definition of a simultaneous mass-matching function 
$\mathcal{G}_{6 \to 4}$:
\begin{equation}
\ln\frac{m_{\ell 6}}{m_{\ell 4}} =
 \int_{\alpha_4}^{\alpha_b}\!dx \frac{\delta_4(x)}{\beta_4(x)}
+ \int_{\alpha_b}^{\alpha_t}\!dx \frac{\delta_5(x)}{\beta_5(x)}
+ \int_{\alpha_t}^{\alpha_6}\!dx \frac{\delta_6(x)}{\beta_6(x)}
+ \mathcal{G}_{6 \to 4}(\alpha_t,\alpha_b) \ .
\label{r4}
\end{equation}
Eqs.~(\ref{r3}) and (\ref{r4}) can be matched to Eqs.~(\ref{p1})--(\ref{p4})
and (\ref{q3}) in the sequential limit (\ref{a1}) to determine
the leading terms in $\mathcal{F}_{6 \to 4}$ (NNLO) and 
$\mathcal{G}_{6 \to 4}$ (NNNLO):
\begin{align} 
\mathcal{F}_{6 \to 4} 
&= - \frac{11}{300\pi}(\alpha_b + \alpha_t) + O(\alpha_{b,t}^2)\ , 
\nonumber \\[1mm]
\mathcal{G}_{6 \to 4} 
&= - \frac{89}{432\pi^2}\bigl(\alpha_b^2 + \alpha_t^2\bigr) 
   - \frac{11}{1725\pi^2}\alpha_b\alpha_t + O(\alpha_{b,t}^3) \ .
\end{align}

For simultaneous $t,b,c$ decoupling, we add
\begin{equation}
\ln(\overline{m}_b/\overline{m}_c)
= \int_{\alpha_c}^{\alpha_b}\!dx/\beta_4(x)
\label{r6}
\end{equation}
to the definitions (\ref{r1}) and specify matching functions 
$\mathcal{F}_{6 \to 3}$ (NNLO) and $\mathcal{G}_{6 \to 3}$ (NNNLO)
as follows:
\begin{align}
\ln\frac{\overline{m}_c}{\bar{\mu}}
&= \int_{\alpha_3}^{\alpha_c}\!dx \frac{1}{\beta_3(x)}
   + \mathcal{F}_{6 \to 3}(\alpha_t,\alpha_b,\alpha_c)  \ ,
 \nonumber \\[1mm]
\ln\frac{m_{\ell 6}}{m_{\ell 3}} &=
  \int_{\alpha_3}^{\alpha_c}\!dx \frac{\delta_3(x)}{\beta_3(x)}
+ \int_{\alpha_c}^{\alpha_b}\!dx \frac{\delta_4(x)}{\beta_4(x)}
+ \int_{\alpha_b}^{\alpha_t}\!dx \frac{\delta_5(x)}{\beta_5(x)} 
+ \int_{\alpha_t}^{\alpha_6}\!dx \frac{\delta_6(x)}{\beta_6(x)}
 \nonumber \\
&\phantom{=\ }
+ \mathcal{G}_{6 \to 3}(\alpha_t,\alpha_b,\alpha_c) \ .
\label{r7}
\end{align}

The dependence of our running couplings on mass logarithms can be 
determined order by order from perturbative results for the 
Callan-Symanzik and matching functions. An example is the LO 
asymptotic behaviour of $\alpha_t$ for simultaneous $t,b,c$ decoupling:
\begin{equation}
\alpha_t \sim 6\pi\bigl/\bigl(23\ln(m_t/\bar{\mu}) 
            + 2\ln(m_b/\bar{\mu}) + 2\ln(m_c/\bar{\mu})\bigr)\,.
\end{equation}
Generally, this coupling runs with a \emph{mixture} of $f$ = 3, 4, and 
5 active flavours.  The sequential limit (\ref{a1}) is a special case, 
where the dominant term in the denominator exhibits the five-flavour 
coefficient 23 ($33 - 2f$ for $f=5$).  Another extreme case is 
$m_t = m_b$ with $\ln(m_c/\bar{\mu})$ neglected, where the coefficient 
becomes 25 ($f=4$). Similarly, we get the $f=3$ coefficient 27 for 
$m_t = m_b = m_c$.

The multi-scale RG can be readily generalized to include logarithmic
momentum scales between the mass logarithms in Eq.~(\ref{eq2}).  The
main new feature is that the invariant masses given by Eqs.~(\ref{r1}) 
and (\ref{r6}) are used to construct scale-invariant logarithms.  For 
example, let $Q$ be a large momentum scale contained in the simultaneous 
limit
\begin{equation}
\ln(m_b/\bar{\mu}) \lesssim \ln(Q/\bar{\mu}) \lesssim \ln(m_t/\bar{\mu})\,
\to\, \mbox{large}.  
\end{equation}
Then amplitudes are expanded as multinomials in $\alpha_t$, $\alpha_b$
and $\alpha_Q$, where $\alpha_Q$ is defined by 
\begin{equation}
\ln(Q/\overline{m}_b) = \int^{\alpha_Q}_{\alpha_b}\!dx/\beta_5(x) 
\end{equation}
or equivalently by
\begin{equation}
\ln(\overline{m}_t/Q) = \int^{\alpha_t}_{\alpha_Q}\!dx/\beta_5(x)\,.
\end{equation}

This completes the formulation of the multi-scale RG.  Its advantage
over the standard method (\ref{eq1}) is that to a given order, no
logarithms are discarded, so uncontrolled approximations or outright
contradictions (such as the example of anomaly cancellation below) 
are avoided.  Each application of the standard method should be 
checked to see how well it approximates the multi-scale result.

The need for a multi-scale approach is best illustrated by the classic
example \cite{CWZ,KM,CK,BCST1,BCST3,Bass} of heavy-quark decoupling from 
the weak neutral axial-vector current 
\begin{equation}
J_{\mu 5}^Z = \frac{1}{2}\biggl\{\sum_{q=u,c,t} - \sum_{q=d,s,b}\biggr\}
                \bigl(\bar{q}\gamma_\mu\gamma_5q\bigr)_6
\end{equation}
where the subscript 6 indicates an operator or amplitude in QCD with 
$f=6$ flavours.  Diagrams for LO contributions in ghost-free gauges
are shown in Fig.~1.

\begin{figure}[t]
\begin{center}
\setlength{\unitlength}{1.1mm}
\begin{picture}(45.6,28)(0,0)
\newcommand{\glue}%
{\curve(0,0, 0.5,0.6, 1,0)\curve(1,0, 1.5,-0.6, 2,0)
\curve(2,0, 2.5,0.6, 3,0)\curve(3,0, 3.5,-0.6, 4,0)
\curve(4,0, 4.5,0.6, 5,0)\curve(5,0, 5.5,-0.6, 6,0)}
\newcommand{\propagator}%
{\thicklines\put(0,0){\glue}
\thinlines
\closecurve(6,0, 8,2, 10,0, 8,-2)
\curve(6,0, 8,2)\curve(6.2,-0.75, 8.7,1.75)
\curve(6.6,-1.4, 9.4,1.4)\curve(7.3,-1.75, 9.8,0.75)\curve(8,-2, 10,0)
\thicklines\put(10,0){\glue}}
\newcommand{\heavy}%
{\thicklines
\put(0,0){\vector(3,2){6}}
\put(0,0){\line(3,2){9}}
\put(9,-6){\vector(-3,2){5.5}}
\put(9,-6){\line(-3,2){9}}
\put(0,0){\circle*{1}}\put(9,-6){\circle*{1}}\put(9,6){\circle*{1}}
\put(9,-6){\line(0,1){12}}
\put(9,6){\vector(0,-1){7}}
\put(10.3,-1){\footnotesize $h$}}
\newcommand{\light}%
{\thicklines
\put(0,-6){\line(3,2){6.6}}
\put(6.6,-1.6){\vector(-3,-2){4.5}}
\put(0,-6){\line(1,0){7.4}}
\put(0,-6){\vector(1,0){5}}
\put(0,6){\line(1,0){7.4}}
\put(7.4,6){\vector(-1,0){5}}
\put(0,6){\line(3,-2){6.6}}
\put(0,6){\vector(3,-2){4.5}}
\put(0,6){\circle*{1}}\put(0,-6){\circle*{1}}
\put(3,-1){\footnotesize $\ell$}}
\newcommand{\proton}
{\thinlines
\put(35.4,24){\line(1,2){3}}\put(35.4,24){\vector(1,2){1.8}}
\put(35.9,23.5){\line(1,2){3.1}}\put(36,23.7){\vector(1,2){1.8}}
\put(36.3,22.8){\line(1,2){3.3}}\put(36.4,23){\vector(1,2){2}}
\put(35.4,8){\line(1,-2){3}}\put(38.4,2){\vector(-1,2){2}}
\put(35.9,8.5){\line(1,-2){3.1}}\put(39,2.3){\vector(-1,2){2}}
\put(36.3,9.2){\line(1,-2){3.3}}\put(39.6,2.6){\vector(-1,2){2}}
\curve(33.6,21.5, 35.5,23.4)
\curve(33.3,19.7, 36.05,22.45)
\curve(33.1,18.0, 36.4,21.3)
\curve(33.0,16.4, 36.65,20.05)
\curve(32.9,14.8, 36.8,18.7)
\curve(33.1,13.5, 36.95,17.35)
\curve(33.3,12.2, 37,15.9)
\curve(33.5,10.9, 36.95,14.35)
\curve(33.9,9.8, 36.75,12.65)
\curve(34.35,8.75, 36.5,10.9)
\linethickness{0.2mm}
\closecurve(33,20, 35,24, 37,20, 37,12, 35,8, 33,12)}
\put(0,13){\small $J_{\mu 5}^Z$}
\put(7,14){\heavy}
\put(16,20){\propagator}\put(16,8){\propagator}
\put(32,14){\light}
\put(6,-2){\proton}
\end{picture}
\end{center}
\caption{Leading-order diagrams for heavy quarks $h$ to decouple from
light quarks $\ell$ when the weak neutral current $J_{\mu 5}^Z$ is
coupled to a nucleon at low momentum transfer. The gluon propagators 
(indicated by wavy lines) are dressed with multiple one-loop self-energy 
insertions.
\label{1}}
\end{figure}
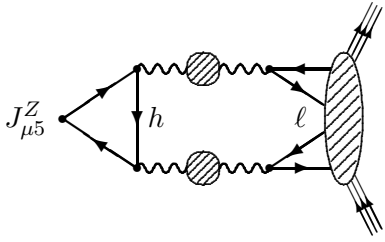

Let us apply the standard procedure (\ref{eq1}) to the six-flavour 
amplitude
\begin{equation}
\Gamma^{(6)}_{\mu 5} = \bigl\langle \bar{t}\gamma_\mu\gamma_5 t
                        - \bar{b}\gamma_\mu\gamma_5 b \bigr\rangle_6\ .
\end{equation}
The first step $m_t \to \infty$ produces a current operator with
scale-invariant renormalization \cite{CK,BCST1}:
\begin{equation}
\lim_{m_t \to \infty} \Gamma^{(6)}_{\mu 5} 
= \Gamma^{(5)}_{\mu 5} = - \bigl\langle 
  (\bar{b}\gamma_\mu\gamma_5 b)_{\mathrm{inv}}\bigr\rangle_5\ .
\end{equation}
In the LO correction to this result,
\begin{equation}
\Gamma^{(6)}_{\mu 5} - \Gamma^{(5)}_{\mu 5} \underset{\mathrm{LO}}{=}
  -\frac{6\run{t}}{23\pi}\Bigl\langle\,\sum_{q=u}^b
 \bigl(\bar{q}\gamma_\mu\gamma_5 q\bigr)_{\mathrm{inv}}\Bigr\rangle_5
\label{t4}
\end{equation}
the Witten running coupling $\run{t}$ displays a \emph{five}-flavour 
dependence \cite{CWZ,CK}:
\begin{equation}
\run{t} \sim \frac{6\pi}{23\ln(m_t/\bar{\mu})}\ .
\end{equation}
This agrees with the rule of thumb that the number of active flavours is
determined by the residual theory.

The problem becomes more subtle when both $t$ and $b$ are to be 
decoupled such that anomaly cancellation occurs for $m_t = m_b$.  The
sequential method (\ref{eq1}) requires us to decouple the
$b$ quark from Eq.~(\ref{t4}),
\begin{equation}
\Gamma^{(6)}_{\mu 5} \underset{\mathrm{LO\;seq}}{=}
\frac{6}{23\pi}\bigl(\run{b} - \run{t}\bigr)\Bigl\langle\,\sum_{q=u}^c
 \bigl(\bar{q}\gamma_\mu\gamma_5 q\bigr)_{\mathrm{inv}}\Bigr\rangle_4
\label{t6}
\end{equation}
but it is immediately evident that the method fails with respect to
cancellation at $m_b = m_t$ because $\run{b}$ runs with \emph{four} 
flavours:
\begin{equation}
\run{b} \sim \frac{6\pi}{25\ln(m_t/\bar{\mu})}\ .
\end{equation}
Presumably this is why the sequential method was not pursued for this
example in \cite{CWZ,CK}; instead, a five-flavour dependence was proposed
for both the $t$ and $b$ contributions:
\begin{equation}
\Gamma^{(6)}_{\mu 5} \overset{\mathrm{?}}{\sim}
\Bigl(\frac{6}{23}\Bigr)^2 \Bigl\{
\frac{1}{\ln(m_b/\bar{\mu})} - \frac{1}{\ln(m_t/\bar{\mu})}\Bigr\}
 \Bigl\langle\,\sum_{q=u}^c
 \bigl(\bar{q}\gamma_\mu\gamma_5 q\bigr)_{\mathrm{inv}}\Bigr\rangle_4\ .
\label{t8}
\end{equation}
Certainly this formula vanishes for $m_t = m_b$, but the five-flavour 
dependence assumed for both scales is not possible: the residual 
theory has only \emph{four} flavours.

The multi-scale RG resolves these problems very simply.  Eq.~(\ref{t6}) 
is a double asymptotic series of the form (\ref{n1}) derived for the
strictly sequential limit (\ref{eq1}).  One cannot apply it directly
to the case $m_t = m_b$ because important LO logarithms have been 
discarded.  To restore all logarithms, replace Witten's running
couplings $\run{t},\run{b}$ by our simultaneous couplings 
$\alpha_t,\alpha_b$ and so obtain a simultaneous expansion of the 
form (\ref{n2}):
\begin{equation}
\Gamma^{(6)}_{\mu 5} \underset{\mathrm{LO\;sim}}{=}
\frac{6}{23\pi}\bigl(\alpha_b - \alpha_t\bigr)\Bigl\langle\,\sum_{q=u}^c
 \bigl(\bar{q}\gamma_\mu\gamma_5 q\bigr)_{\mathrm{inv}}\Bigr\rangle_4\ .
\label{u1}
\end{equation}
Since$^6$
$\alpha_t = \alpha_b$ for $m_t = m_b$,
anomaly cancellation is automatic; at the same time, this result
includes the sequential formula~(\ref{t6}) as a special case. These 
properties are especially evident in the asymptotic result replacing 
(\ref{t8}):
\begin{equation}
\Gamma^{(6)}_{\mu 5} \underset{\mathrm{sim}}{\sim}
   \frac{36}{23}\Bigl[\Bigl\{25\ln\frac{m_b}{\bar{\mu}}\Bigr\}^{-1}
           - \Bigl\{23\ln\frac{m_t}{\bar{\mu}} 
            + 2\ln\frac{m_b}{\bar{\mu}}\Bigr\}^{-1}\Bigr]
 \Bigl\langle\,\sum_{q=u}^c
 \bigl(\bar{q}\gamma_\mu\gamma_5 q\bigr)_{\mathrm{inv}}\Bigr\rangle_4\ .
\label{u2}
\end{equation}
A direct analysis of the diagrams in Fig.~\ref{1} gives the same 
answer \cite{BCST3}.

The next order in the simultaneous expansion can be deduced from the 
NLO correction to the sequential result (\ref{t6}) found in
\cite{BCST1}.  Making the replacement $\run{t,b} \to \alpha_{t,b}$,
we find:
\begin{equation}
\Gamma^{(6)}_{\mu 5} \underset{\mathrm{NLO}}{=} 
 \frac{6}{23\pi}\bigl(\alpha_b-\alpha_t\bigr)  
 \Bigl\{1 + \frac{125663}{82800\pi}\alpha_b
          + \frac{6167}{3312\pi}\alpha_t\Bigr\}
 \Bigl\langle\,\sum_{q=u}^c
 \bigl(\bar{q}\gamma_\mu\gamma_5 q\bigr)_{\mathrm{inv}}\Bigr\rangle_4\ .
\end{equation}
The factor $\alpha_b-\alpha_t$ ensures that anomaly cancellation occurs
automatically at $m_t = m_b$.

In practice, the multi-scale RG will be important in determining
momentum dependence between heavy-quark thresholds, e.g.\ in deep inelastic
scattering.  It will allow the whole range of momenta, from one threshold
to the next, to be described by a single expansion calculable (in principle)
to any order.

When a threshold is crossed, the multi-scale asymptotic expansion changes
because the relevant operator-product and heavy-quark expansions 
\emph{differ}. Consequently there are slope discontinuities at 
thresholds, as for other mass-independent methods \cite{marc}.  The 
standard prescription for the removal of discontinuities is the method of 
effective charges \cite{brod,grun,kenn,shir}, but this has been 
set up only for the single-scale RG.  Formulating a natural multi-scale
extension of this method presents an interesting challenge for the future.

The result of this letter is a substantial generalization of the RG
which solves problems involving logarithms for more than one large 
scale.  Amplitudes are expanded as multinomials of running couplings, 
one for each large scale but depending (in general) on all of these 
scales; this replaces the familiar perturbation series in one running
coupling for a single-scale RG.  All multi-scale results to a given order 
(degree in the multinomial expansion) can be readily deduced from 
single-scale calculations to the same order.  Our method avoids 
inconsistencies by retaining \emph{all} important logarithms to a
given order.


\end{document}